\documentclass[showpacs,preprintnumbers,amsmath,amssymb,superscriptaddress]{revtex4}

\usepackage{graphicx}
\usepackage{dcolumn}
\usepackage{bm}

\begin{document}

\title{Entropy radiated by a braneworld black hole
}
\author{Aur\'elien Barrau}
\email{Aurelien.Barrau@cern.ch}
\affiliation{Laboratoire de Physique Subatomique et de Cosmologie,\\
 UJF, CNRS/IN2P3, INPG\\
53 avenue des Martyrs, 38026 Grenoble cedex, France}
\author{Julien Grain}
\email{grain@apc.univ-paris7.fr}
\affiliation{AstroParticule et Cosmologie, Universit\'e Paris 7, CNRS/IN2P3 \\
10, rue Alice Domon et L\'eonie Duquet, 75205 Paris cedex 13, France}
\author{Carole Weydert}
\email{weydert@lpsc.in2p3.fr}
\affiliation{Laboratoire de Physique Subatomique et de Cosmologie,\\
UJF, CNRS/IN2P3, INPG\\
53 avenue des Martyrs, 38026 Grenoble cedex, France}


\date{\today}

\begin{abstract}
The concept of black hole entropy is one of the most important enigmas of 
theoretical physics. It relates thermodynamics to gravity and allows substantial
hints toward a quantum theory of gravitation. Although Bekenstein conjecture
--assuming the black hole entropy to be a measure of the number of precollapse 
configurations-- has proved to be extremely fruitful, a direct and conclusive
proof is still missing. This article computes accurately the entropy evaporated by
black holes in $(4+n)$ dimensions taking into account the exact greybody factors.
This is a key process to constrain and understand the entropy of black holes as 
the final state is unambiguously defined. Those results allow to generalize
Zurek's important argument, in favor of the Bekenstein conjecture, to 
multi-dimensional scenarios.
\end{abstract}

\maketitle

This article aims at generalizing to an arbitrary number of dimensions the key argument on the entropy 
of black holes made 
by Zurek \cite{zurek}, improved by Page \cite{page1} and reconsidered by
Schumacher \cite{schumacher}. This is a mandatory step to consider black holes
in braneworld scenarios. As it requires quite sophisticated numerical techniques
to compute the greybody factors, this point could not have been easily made
earlier.

Bekenstein \cite{bek1} proposed that black holes possess an entropy proportional
to their surface area, which can be statistically interpreted as a measure of the
number of precollapse configurations leading to a given state. It can easily be
shown that the only possible power-law function of the area $A$ allowed for the
entropy is linear (see, {\it e.g.} \cite{bek2}, for an introductory review). The
constant of proportionality can be fixed using the Hawking law \cite{hawking},
leading to $S=A/4$ in Planck units. Attempts to understand this simple formula
from more fundamental points of view are legion. Important historical points
were made by Bombelli {\it et al.} \cite{bomb} and Srednicki \cite{sre} to
relate the entropy of black holes to entanglement entropy arising from the tracing out
of those degrees of freedom localized beyond the horizon. Thorne and Zurek
\cite{thorn} and 't Hooft \cite{hoo} considered the thermal radiative
"atmosphere" of black holes. More recently, the entropy has been linked with the
degrees of freedom of strings associated with the hole \cite{stro,mal} or with the
quantum gravity degrees of freedom of the horizon, in the context of
conformal field theory \cite{car,solo} or in loop quantum gravity \cite{ash}. More
heuristic schemes were also developed, {\it e.g.} in \cite{muk}. Basically
speaking, recovering the Bekenstein black hole entropy is one of the most
important successes of the main tentative theories of quantum gravity. This
happens quite naturally in string theory (at least for {\it some} extremal black holes --
see \cite{dam,dav} for reviews) and in loop quantum gravity (at least when 
setting correctly the free Imirzi parameter -- see \cite{ash2}).

However, a direct proof of Bekenstein's interpretation of the entropy (relating its
meaning with the outer world) is still
missing. Zurek's idea to go further in this direction was to consider the black 
hole evaporation and to quantify the ratio of the evaporated entropy to the
entropy lost by the black hole \cite{zurek}. This is a fundamental process to
this purpose because the final state is well determined and its entropy can be
accurately estimated. The key ingredient for this
computation --namely the greybody factors-- were derived by Page \cite{page2} (following
Starobinsky \cite{staro1}), 
thanks to exhaustive numerical computations,
allowing to solve this problem exactly in 4 dimensions \cite{page2,page1}.
However, quite a lot of modern fundamental models of theoretical physics 
--beginning of course with string theory-- involve extra-dimensions. We now 
proceed to the
investigation of this problem in the Arkani-Hamed-Dimopoulos-Dvali (ADD)
scenario \cite{ADD} based on large extra-dimensions which accounts, among other
problems, for the huge hierarchy between the different scales of physics. The
keypoint is to allow only gravitons to propagate in the extra dimensions while
standard model fields are localized on the brane.\\

The metric for a $(4+n)$-dimensional black holes reads as 
$$
ds^2=-h(r)dt^2+h(r)^{-1}dr^2+r^2 d\Omega^2_{n+2},
$$
where
$
d\Omega_{n+2}$ is the solid angle element
and
$
h(r)=1-\left(r_H/r\right)^{n+1},
$
with $r_H$ the horizon radius. The temperature $T$ of such black holes is given by
$$
T = \left.\frac{1}{4\pi}\frac{dh}{dr}\right|_{r_H}= \frac{n+1}{4\pi r_H},
$$
the radius being related with the mass $M$ of the hole through
$$
r_H=\frac{1}{\sqrt{\pi}M_{*}}\left(\frac{M}{M_{*}}\right)^{\frac{1}{n+1}}\left(\frac{8}{n+2}\right)^{\frac{1}{n+1}}\left[\Gamma\left(\frac{n+3}{2}\right)\right]^{\frac{1}{n+1}},
$$
where $M_*$ is the fundamental $(4+n)$-dimensional Planck scale, while the area is given by
$$
A=r_H^{n+2} (2\pi) \pi^{\frac{n+1}{2}}\left[\Gamma\left(\frac{n+3}{2}\right)\right]^{-1}.
$$
As a first approximation, for particles confined on the brane, the radiation emitted in a short
time interval $dt$ carries an energy $dE=\Sigma \alpha T^{4}dt$ and an entropy 
$dS_{rad}\approx\frac{4}{3}\Sigma \alpha T^{3} dt$ where $\Sigma$ is an effective cross section
and $\alpha$ is the Stefan constant. This allows to compute the variation of the black-hole
entropy:
\begin{eqnarray}
dS_{BH} & = & \frac{dA}{4}M_{*}^{n+2} =\frac{M_{*}^{n+2}}{4}\left[A(M)-A(M-dE)\right]\nonumber \\
& \approx & f(n) \frac{n+2}{n+1}M^{\frac{1}{n+1}}dE\nonumber\label{Sbh}
\end{eqnarray}
with
$$
f(n)=\frac{\sqrt{\pi}}{2}\frac{1}{M_{*}^{\frac{n+2}{n+1}}}\left(\frac{8}{n+2}\right)^{\frac{n+2}{n+1}}\left[\Gamma\left(\frac{n+3}{2}\right)\right]^{\frac{1}{n+1}}.
$$
The ratio $R$ of the entropy lost by the black hole to the entropy gained by
the radiation can be estimated:
$$
R=\frac{dS_{rad}}{dS_{BH}}=\frac{4}{3}\frac{n+1}{n+2}\frac{1}{f(n)T}M^{-\frac{1}{n+1}}=\frac{4}{3}.
$$
This is exactly the same result as obtained by Zurek in 4 dimensions. As one could expect, the 
effect of the modification of the area of the black hole due to the extra-dimensions is totally 
compensated by the modification of its temperature. However, it will be demonstrated that a more 
rigorous treatment of this problem radically modifies this conclusion. This is due both to the 
emission of gravitons in the bulk and to the footprint of the extra-dimensions on the greybody
factors, even for brane fields.\\

\begin{figure}[ht]
	\begin{center}
		\includegraphics[scale=0.45]{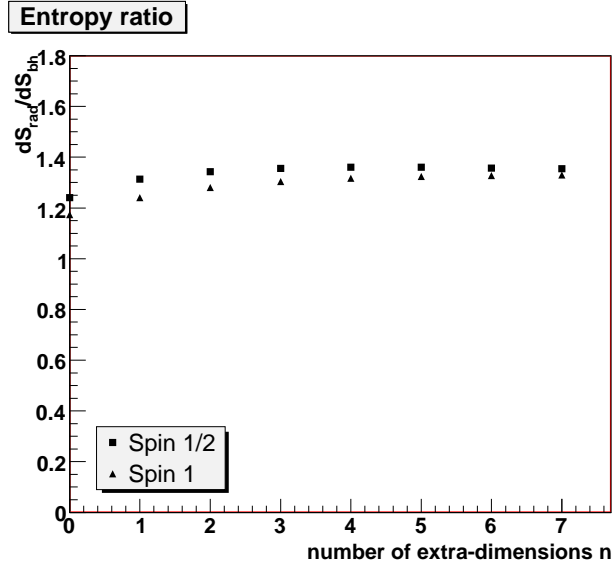}
	\end{center}
	\caption{"Macroscopic" ratio of the radiated entropy to the variation of the hole entropy, as
	a function of the number of extra-dimensions for fermions and gauge bosons confined on the
	4-brane.}
	\label{rap1}
\end{figure}

The exact semi-classical flux spectrum of a higher-dimensional black hole is given by:
$$
\frac{dN^{(s)}(\omega)}{dt}=\sum_{\ell,m}^{}\sigma^{(s)}_n(\omega)\frac{1}{\exp(\frac{\omega}{T})-(-1)^{2s}}
\frac{d^{n+3}k}{(2\pi)^{n+3}}
$$
where the cross section $\sigma$ accounts for the couplings between the black hole and quantum
fields with spin $s$. Using the "macroscopic" thermodynamical approach of Zurek, one can evaluate $R$ for
standard model fields through
$$
R=\frac{dS_{rad}}{dS_{BH}}=\frac{\int_{0}^{\infty}\sigma
\left[\frac{x}{e^x-1}-\ln(1-e^{-x})\right]x^2\, dx}{\int_{0}^{\infty}\sigma\frac{x^3}{e^x-1}\,
dx}.
$$
The results are given, as a function of the number of extra-dimensions, in Fig.~\ref{rap1}. A weak dependance
upon $n$ can be noticed, together with a good agreement with the black-body approximation.

\begin{figure}[ht]
	\begin{center}
	\vspace{0.45cm}
		\includegraphics[scale=0.65]{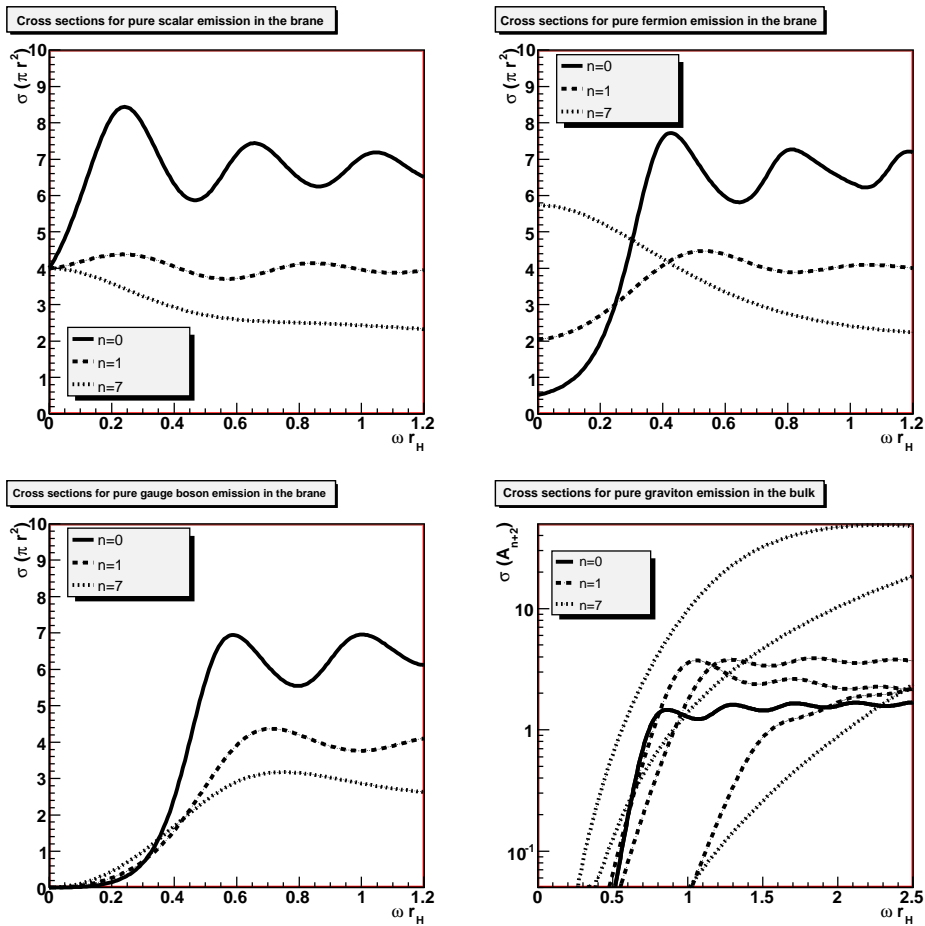}
	\end{center}
	\caption{Cross-sections as a function of the energy of the particle multiplied by the
	horizon radius for n=0, n=1 and n=7. Upper-left: scalars on the brane, upper right:
	fermions on the brane, lower left: gauge bosons on the brane, lower right: gravitons
	in the bulk. For this latter plot the three curves correspond (respectively from top to
	bottom at $\omega r_h=2$) to the tensor, vector and scalar contributions.}
	\label{grey}
\end{figure}

Although more meaningful than the first approximation, this approach still misses the fundamental point, 
{\it i.e.} an accurate account for the greybody factors when the entropy is correctly expressed 
with the density matrix as $tr(-\rho ln\rho)$.
For a non-rotating, uncharged, black hole  the
radiation is emitted with a density matrix in each mode 
$$\rho_{kk'}=\delta_{kk'}\frac{\left|A_{\ell,s}\right|^{2k}(e^{\frac{\omega}{T}}-(-1)^{2s})}{(e^{\frac{\omega}{T}}-(-1)^{2s}+(-1)^{2s}\left|A_{\ell,s}\right|^{2})^{k+(-1)^{2s}}}$$
where $\left|A_{\ell,s}\right|^2$ is the transmission coefficient of the black hole to the considered particle.
This allows to write
$$
\frac{dS_{BH}}{dt}=-\frac{1}{T}\frac{dM}{dt}=-\frac{1}{T}\displaystyle\sum_{\ell,m,s}\int\frac{d\omega}{2\pi}\omega\underbrace{\frac{\left|A_{\ell,s}\right|^2}{e^{\omega/T}-(-1)^{2s}}}_{\left<N_{\omega,\ell,m}\right>},
$$
where the summation over the angular momentum $\ell$ and its projection $m$ must be carried out
after performing the integrals, and:
\begin{widetext} 
$$
\frac{dS_{rad}}{dt}=\displaystyle\sum_{\ell,m,s}\int\frac{d\omega}{2\pi}\left[\underbrace{\frac{\left|A_{\ell,s}\right|^2}{e^{\omega/T}-(-1)^{2s}}\ln{\left(\frac{e^{\omega/T}-(-1)^{2s}}{\left|A_{\ell,s}\right|^2}+(-1)^{2s}\right)}+(-1)^{2s}\ln{\left(1+(-1)^{2s}\frac{\left|A_{\ell,s}\right|^2}{e^{\omega/T}-(-1)^{2s}}\right)}}_{\left<S_{\omega,\ell,m}\right>}\right].
$$
\end{widetext} 
The coefficients $A_{\ell,s}$ account both for the intricate metric structure around the black hole
and for the centrifugal barrier, possibly leading to a back-scattering of the emitted particle
\cite{kanti}. They can be determined by solving the equations of motion with
appropriate boundary conditions at the black hole's horizon and at spatial infinity. 
For particles confined on the brane --with the usual anzatz $\Phi_s(t,r,\Omega)=e^{-i\omega{t}}Y^\ell_{m,s}(\Omega)R_s(r)$ to split the
temporal, angular and radial parts of the field (where $Y^\ell_{m,s}$ are the spin-weighted spherical
harmonics)-- the radial function $R_s(r)$ obeys the master equation:
$$
\left[\Delta^s\frac{d}{dr}\left(\Delta^{1-s}\frac{d}{dr}\right)+
\left(\frac{\omega^2r^2}{h}+2is\omega{r}-\frac{is\omega{r}^2}{h}\frac{dh}{dr}-\lambda\right)\right]P_s=0
$$
where $\Delta=r^2h(r)$, $P_s(r)=\Delta^sR_s(r)$ and $\lambda=j(j+1)-s(s-1)$. This is equivalent,
in the tortoise coordinate system,                 
to a Shr\"odinger equation with a potential depending on the metric and its derivatives, on the
number of extra-dimensions and on the spin of the field.
The keypoint is that the derived greybody factors keep a clear footprint of the
dimensionality of space, even for those particles living on the 4-brane. This is the main reason why
the entropy ratios are strongly modified by the existence of extra-dimensions. The case of
gravitons --propagating in the bulk-- is more subtle, requiring to investigate the three kind
(scalar, vector, tensor) of gravitational perturbations evolving in a higher dimensional
space (see, {\it e.g.}, Ref. \cite{kodama,creek,cardoso}).
Although analytical
expressions can be obtained in the infra-red and ultra-violet limits for all the aforementionned cases, the most interesting regime,
where $\omega r_H \sim 1$, requires dedicated numerical investigations for each kind of particle. We have solved
the appropriate field equations with an extensive simulation (see, {\it e.g.}, \cite{grain} for a
description) which, basically, proceeds in three steps: for each value of the energy $\omega$
and of the angular momentum $\ell$, the radial part of the equations is numerically solved, then the
amplitudes of the outgoing and ingoing modes at spatial infinity are estimated by fitting the
asymptotic analytical solutions and, finally, the resulting transmission coefficient is used to compute the
contribution to the $\ell$-th multipole. Some examples for 0, 1 and 7 extra-dimensions are given
in Fig.~\ref{grey}.

Using those transmission coefficients, it is possible to compute exactly the entropy lost by the
black hole and gained by the radiation. Results are given in Tab.~\ref{res}, as a function of the
number of degrees of freedom, and constitute the first computation of the entropy radiated by a 
braneworld black hole. This is the key result of this paper. Remarkably, this process allows to relate the --speculative-- entropy of 
the black hole in $(4+n)$ dimensions with the well known entropy of a multidimensional blackbody radiation.
Throughout this paper the emitted particles are
assumed to be massless. Although taking into account the mass will inevitably
modify the greybody factors \cite{grain2}, this will not change our conclusions
as the point remains valid for most black hole temperatures. When
the black hole is very massive ($T\ll \mu_{min}$ where $\mu_{min}$ is the mass of
the lightest particle) only massless fields will be radiated and when the black
hole is very light  ($T\gg \mu_{max}$ where $\mu_{max}$ is the mass of
the heaviest particle) ultra-relativistic quanta will be radiated. Important modifications
of the transmission coefficients are therefore only expected when $T\in[\mu_{min},\mu_{max}]$.

\begin{figure}[ht]
	\begin{center}
		\includegraphics[scale=0.45]{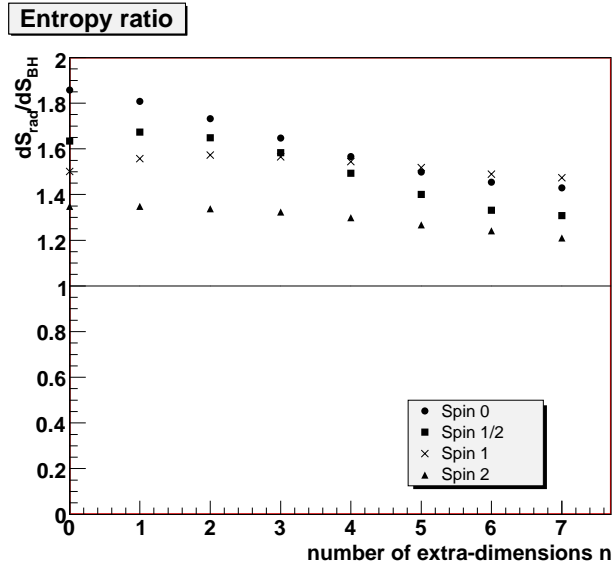}
	\end{center}
	\caption{Exact computation of the ratio of the radiated entropy to the variation of the
	black hole entropy, as
	function of the number of extra-dimensions for scalars, fermions and gauge bosons confined on the
	4-brane and gravitons propagating in the bulk.}
	\label{rap}
\end{figure}

The ratios $R=dS_{rad}/dS_{BH}$ are plotted in Fig.~\ref{rap}. It should be noticed 
that, contrarily to what happens in Zurek's
approach, nothing here mathematically prevents $R$ from being arbitrarily large or small. The fact
that the values computed are very close to one is a strong point in favor of the conjecture of
Bekenstein. Furthermore, it is still possible --at least at the {\it gendankenexperiment} level-- 
to consider a black hole in thermal equilibrium with a surrounding heat bath. The condition of
equilibrium simply requires the energy associated with the radiation to be much smaller than the
mass of the hole \cite{hawking2}. This straightforwardly generalizes to an arbitrary
number of dimensions. In this situation, 
adiabatically changing the volume of the container so that the entropy production ratio remains
arbitrarily close to one allows to reinforce the validity of the generalized second law of thermodynamics.

Finally, one could also play the game to consider a black hole emitting all the fields in all the
extra-dimensions. An elementary thermodynamical consideration allows to prove that, at the first
order of approximation, $R=\frac{n+4}{n+3}$. It means that, as the number of dimensions increases,
the evaporation process becomes more and more isentropic. Or, the other way round, that the
subtle correlations assumed in 4 dimensions to account for the "artificial" entropy increase
during the evaporation process become less and less important. This would mean that the coarse-grained entropy
becomes equivalent to the fine-grained entropy, therefore solving the unitarity problem.
A more refined numerical study shows
that this estimate of $R$ is correct at the percent level.

\begin{widetext}
\begin{center}
\begin{table}[ht]
\begin{tabular}{|c||c|c|}
\hline n & $dS_{rad}/dt$ $(10^{-3}r_{H}^{-1})$ & $dS_{BH}/dt$
 $(10^{-3}r_{H}^{-1})$ \\ \hline
\hline 0 &  $(6.948\; n_{0}+3.368\; n_{1/2}+ 1.269\; n_{1}+0.260\;n_2)$
 & $(3.740\; n_{0}+2.060\; n_{1/2}+ 0.846\; n_{1}+0.193\;n_2)$ \\  
\hline 1 &  $(30.32\; n_{0}+24.39\; n_{1/2}+ 17.87\; n_{1}+13.10\;n_2)$
 & $(16.77\; n_{0}+14.57\; n_{1/2}+ 11.48\; n_{1}+9.712\;n_2)$ \\  
\hline 2 &  $(79.12\; n_{0}+68.35\; n_{1/2}+ 64.19\; n_{1}+88.75\;n_2)$
 & $(45.71\; n_{0}+41.44\; n_{1/2}+ 40.80\; n_{1}+66.33\;n_2)$ \\  
\hline 3 &  $(160.7\; n_{0}+139.8\; n_{1/2}+ 147.5\; n_{1}+326.3\;n_2)$
 & $(97.57\; n_{0}+88.34\; n_{1/2}+ 94.27\; n_{1}+246.4\;n_2)$ \\  
\hline 4 &  $(280.8\; n_{0}+245.4\; n_{1/2}+ 273.1\; n_{1}+988.2\;n_2)$
 & $(179.3\; n_{0}+164.4\; n_{1/2}+ 176.9\; n_{1}+760.8\;n_2)$ \\  
\hline 5 &  $(445.4\; n_{0}+394.2\; n_{1/2}+ 446.1\; n_{1}+2920 \;n_2)$
 & $(297.2\; n_{0}+281.3\; n_{1/2}+ 294.1\; n_{1}+2305 \;n_2)$ \\  
\hline 6 &  $(660.7\; n_{0}+595.6\; n_{1/2}+ 671.0\; n_{1}+8881 \;n_2)$
 & $(457.1\; n_{0}+447.4\; n_{1/2}+ 449.9\; n_{1}+7183 \;n_2)$ \\  
\hline 7 &  $(932.3\; n_{0}+846.0\; n_{1/2}+ 950.7\; n_{1}+28386\;n_2)$
 & $(661.3\; n_{0}+646.5\; n_{1/2}+ 645.0\; n_{1}+23466\;n_2)$ \\
  \hline  
\end{tabular}
\caption{Rates of entropy variation for the radiation and for the black hole, in units of $10^{-3}r_{H}^{-1}$ as a function of the
number of extra-dimensions $n$ and of the scalar ($n_0$), fermionic ($n_{1/2}$), bosonic ($n_1$)
and gravitational ($n_{2}$) degrees of freedom. Gravitons are assumed to propagate in the bulk whereas the other
fields propagate in the brane.}
\label{res}
\end{table}
\end{center}
\end{widetext}

This article provides the first computation of the entropy evaporated by a braneworld black hole.
The results are given at three different levels of approximation, the latter taking into account
the complete numerical evaluation of the greybody factors. This relates entropies 
of totally different conceptual origins, one being gravitational --associated with 
the area of the black hole-- and the other being statistical --associated with the 
emitted radiation. This is a new step toward a full understanding of the entropy 
of black holes in mutli-dimensional scenarii required by most particle physics and 
cosmology unification models. Remarkably, it does not rely on an assumed 
fundamental theory ({\it e.g.} strings or loops) but just on the well established 
semi-classical evaporation of black holes. This favours
the conjecture of Bekenstein in (4+n)-dimensions and, most importantly, allows
exact computations of the entropy ratios.

\end{document}